\documentstyle[12pt]{article}
\begin{document}
\parskip15pt
\def\complex{{\kern .1em {\raise .47ex \hbox {$\scriptscriptstyle |$}}
\kern -.4em {\rm C}}}
\countdef\refno=30
\refno=0
\def\ref{\advance \refno by 1 \ifnum\refno<10 \item{ [\the\refno]} \else
\item{[\the\refno ]} \fi}
\hfuzz=15pt
\null
\vskip 3cm
{\centerline{{\bf REPRESENTATION THEORY OF QUANTIZED POINCARE
ALGEBRA.}}
\centerline{{\bf TENSOR OPERATORS AND THEIR APPLICATIONS}}
\centerline{{\bf TO ONE-PARTICLE SYSTEMS}}
\bigskip
\centerline{H. Ruegg \footnote
{E-mail~:
$\langle$RUEGG at sc2a.unige.ch$\rangle$.} \footnote
{Partially
supported by the Swiss National Science Foundation.}
and V.N. Tolstoy$^{2}$
\footnote
{Partially supported by a
Soros Foundation Grant of the American
Physical Society.} \footnote
{Permanent Address~:
Institute of Nuclear Physics, Moscow
State University, 119899 Moscow, Russia. E-mail~: $\langle$TOLSTOY at
compnet.msu.su$\rangle$.}}
\centerline{D\'epartement de Physique Th\'eorique}
\centerline{Universit\'e de Gen\`eve}
\centerline{CH-1211 Gen\`eve 4, Switzerland}
\vskip 1cm
\centerline{UGVA-DPT 1993/07-828}
\null
\vskip 1cm
\centerline{ABSTRACT}
\medskip
A representation theory of the quantized Poincar\'e ($\kappa$-Poincar\'e)
algebra (QPA) is developed. We show that the representations of
this algebra are closely connected with the representations of
the non-deformed Poincar\'e algebra. A theory of tensor operators
for QPA is considered in detail. Necessary and sufficient conditions
are found in order for scalars to be invariants.
Covariant components of the four-momenta and the Pauli-Lubanski
vector are explicitly constructed.These results are used for
the construction of some q-relativistic equations.
The Wigner-Eckart theorem for QPA is proven.
\vfill\eject

\noindent
1. {INTRODUCTION}

Recently a quantum deformation of nonsemisimple Lie algebras has
been proposed, using the contraction procedure [1].
This led in particular to the quantum deformation of the
universal enveloping algebra $U({\cal P})$ of the four-dimensional
Poincar\'e algebra [2], [3], [4], which we denoted by $U_q ({\cal P})$
or $U_{\kappa} ({\cal P})$ ($q$ being replaced by the dimensionful
$\kappa$). This result has been generalized to any dimension [5].
In this way one obtains real noncocommutative $*$-Hopf algebras
[4,5].

This new invariance algebra is a non ad hoc way of introducing a
minimal length or a minimal time scale into relativistic theories.
The quantum deformation implies deformed Casimir operators (i.e.
dispersion relations) and deformed wave equations (Klein-Gordon,
Dirac equations, etc).

Comparison with experiment yields limits on a possible nonlocality
[6,7,8,9].
On the other hand this nonlocality can serve as a cutoff in field
theory.

Several steps have already been taken toward a representation theory
of $U_{\kappa} ({\cal P})$ [3,4,7], [10,11].
In this paper we intend to make a systematic study of this important
question.
The tool which we propose to use takes full advantage of the Hopf
algebra structure.
This tool is the quantum adjoint action, which is a homomorphism of
the universal enveloping algebra $U_{\kappa} ({\cal P})$.

The plan of the paper is the following~: In Section 2 we recall the
Hopf algebra structure of $U_q ({\cal P})$ (commutation relations,
coproducts, counits, antipodes).
In Section 3 we discuss the general relations between the
representation of the quantized and non-quantized Poincar\'e algebra.
In Section 4 the main properties of the adjoint action are given.
Section 5 compares scalars and invariants in the quantized algebra.
Section 6 gives the explicit form of the irreducible tensor operators
of $U_q ({\cal P} )$.
In Section 7 the quantum deformed relativistic equations are constructed.
The proof of the Wigner-Eckart theorem is given in section 8, followed
by the conclusion.

\noindent
2. {THE QUANTIZED POINCARE ALGEBRA (QPA)}

The quantum deformation of the universal enveloping algebra
$U({\cal P})$ of the Poincar\'e algebra ${\cal P}$ is an unital
associative algebra $U_q ({\cal P})$ with the following generators~:
the angular moments $J_i$, $i = 1,2,3$, the Lorentz boosts $N_i$,
$i = 1,2,3$, and the four-momenta $P_i$, $P_0$, $i = 1,2,3$, which
satisfy the following relations \footnote
{Everywhere we use the following
standard conventions of relativistic physics~: repeated indices
mean a summation; latin indices run over $1,2,3$, and greek ones run
over $0,1,2,3$.}
($j,k,l = 1,2,3$; $\mu ,\nu = 0,1,2,3)$~:
$$
[J_j ,J_k ] = i\varepsilon_{jkl} J_l \ ,
\eqno {\rm (2.1a)}
$$
$$
[J_j ,N_k ] = i\varepsilon_{jkl} N_l \ ,
\eqno {\rm (2.1b)}
$$
$$
[J_j ,P_k ] = i\varepsilon_{jkl} P_l \ ,
\eqno {\rm (2.1c)}
$$
$$
[J_j ,P_0 ] = 0 \ ,
\eqno {\rm (2.1d)}
$$
$$
[N_j ,N_k ] = -i\varepsilon_{jkl} \{ {1 \over 2} J_l (q^{P_0} +
q^{-P_0} ) - {1 \over 4} (\ln q)^{2} P_l (\vec P \vec J) \} \ ,
\eqno {\rm (2.1e)}
$$
$$
[N_j ,P_k ] = {i \over 2} \delta_{jk} (\ln q)^{-1} (q^{P_0} -
q^{-P_0} )\ ,
\eqno {\rm (2.1f)}
$$
$$
[N_j, P_0 ] = iP_j \ ,
\eqno {\rm (2.1g)}
$$
$$
[P_{\mu}, P_{\nu} ] = 0 \ ,
\eqno {\rm (2.1h)}
$$
where the parameter $q$ is given by
$$
q := \exp {1 \over \kappa} \ ,
\eqno {\rm (2.1i)}
$$
and $\varepsilon_{jlk}$ is the totally antisymmetric tensor
$(\varepsilon_{123} = 1)$; $\delta_{jk}$ is the Kronecker symbol;
$\kappa$ is a parameter with the same dimension as $P_0$.

The standard Hopf structure of $U_q ({\cal P})$ is defined by the
following formulas for a coproduct $\Delta_{q'}$, an antipode
$S_{q'}$ and a counit $\varepsilon := \varepsilon_{q'}$~:
$$
\Delta_{q'} (J_j ) = J_j \otimes 1 + 1 \otimes J_j \ ,
\eqno {\rm (2.2a)}
$$
$$
\Delta_{q'} (N_j ) = N_j \otimes q'^{{P_0 \over 2}} +
q'^{-{P_0 \over 2}} \otimes N_j +
{1 \over 2} (\ln q')\varepsilon_{jlk} (P_l \otimes J_k
q'^{{P_0 \over 2}} + q'^{-{P_0 \over 2}} J_l \otimes P_k ) \ ,
\eqno {\rm (2.2b)}
$$
$$
\Delta_{q'} (P_j ) = P_j \otimes q'^{{P_0 \over 2}} +
q'^{-{P_0 \over 2}} \otimes P_j \ ,
\eqno {\rm (2.2c)}
$$
$$
\Delta_{q'} (P_0 ) = P_0 \otimes 1 + 1 \otimes P_0 \ ,
\eqno {\rm (2.2d)}
$$
$$
S_{q'} (J_j ) = -J_j \ ,
\eqno {\rm (2.3a)}
$$
$$
S_{q'} (N_j ) = -N_j + {3i \over 2} (\ln q')P_j \ ,
\eqno {\rm (2.3b)}
$$
$$
S_{q'} (P_{\mu}) = -P_{\mu} \ ,
\eqno {\rm (2.3c)}
$$
$$
\varepsilon (J_j ) = \varepsilon (N_j ) = \varepsilon (P_{\mu}) =
0\ ,\ \varepsilon (1) = 1 \ ,
\eqno (2.4)
$$
where $j = 1,2,3$; $\mu = 0,1,2,3$; $q' = q$ or $q^{-1}$.
Thus there are two standard Hopf structures for $q' = q$ and
$q' = \bar q := q^{-1}$.
It should be noticed that
$$
\Delta'_{q'} = \Delta_{\bar q'}
\eqno (2.5)
$$
where by $\Delta'$ we denote the opposite coproduct, i.e. if
$\Delta (a) = \sum_i a_i \otimes b_i$ then
$\Delta' (a) = \sum_i b_i \otimes a_i \ .$

\noindent
{\it Remarks}. (i) The defining relations (2.1a-h) are invariant
with respect to the exchange $q \to q^{-1}$. \hfill\break
(ii) If the parameter $\kappa$ in (2.1i) lives on the real axis then
this deformation is called of hyperbolical type and if $\kappa$
lives on the imaginary axis then this deformation is called of
spherical type.
The first type was obtained in [4] and the second one in [2]
by the Wigner-In\"on\"u  contraction of some real forms of the quantum
algebra $U_q (B_2 )$ $(U_q (so(5,\complex ))$ . \hfill\break
(iii) The Hopf algebra $(U_q ({\cal P}), \Delta_q ,S_q ,\varepsilon )$
is called also the quantized Poincar\'e algebra ${\cal P}$ (QPA).
This terminology will be used by us too.

\noindent
3. RELATION BETWEEN REPRESENTATIONS OF THE QUANTIZED AND
\hfill\break
NON-QUANTIZED POINCARE ALGEBRAS

As is well known, the representation theory of the quantized simple Lie
algebras (for generic $q$) is closely connected with the representation
theory of the non-quantized case. Namely, a theorem states that any
irreducible representation (IR) (from the category {\cal O})
of any non-quantized simple Lie algebra can be lifted to an IR of
the corresponding quantized algebra. And also any IR (from the category
{\cal O}) of any quantized simple Lie algebra (for generic $q$)
is converted, when $q \to 1$, to an IR of a corresponding non-quantized
Lie algebra.

An analogous theorem exists also for the quantized Poincar\'e algebra.
Moreover, for the usual Poincar\'e algebra there is a simple and very
important class of IR's which are simultaneously IR's of
$U_q({\cal P})$.
We start to describe some properties of $U_q({\cal P})$ IR's from
this simple class.

\noindent
PROPOSITION 3.1 . {\it Any representation of $U_q ({\cal P})$, where the
operators of the four-momenta act trivially (i.e. as zero operators)
coincides with a representation of the classical Poincar\'e algebra
${\cal P}$.}

\noindent
{\it Proof}. The proposition affirms that when the four-momenta
operators are equal to zero, $P_{\mu} = 0$, then the operators
$J_j $, $ N_j $, $j = 1,2,3$, have the same commutation relations
as the corresponding operators of the usual Lorentz algebra.
Indeed, putting $P_{\mu} = 0$ in the relations (2.1a-h) we see
this to be the case.

\noindent
COROLLARY. {\it Any representation of the Lorentz algebra
is trivially extended to a representation of ${\cal P}$ and
$U_q ({\cal P})$ if we put $P_{\mu} = 0$, $\mu = 0,1,2,3$.}

This simple proposition is very important from the practical point
of view because it allows us to use for such types of representations
of $U_q({\cal P})$ the usual Clebsch-Gordan coefficients of the Lorentz
algebra.

Now we formulate a general theorem about a relation between Hermitian
representations of the quantized and non-quantized Poincar\'e algebras.
We recall that a Hermitian IR of the quantized Poincar\'e algebra is
(like in the non quantized case) characterized by the the eigenvalues of
the two Casimir operators $C_1 (q)$ and $C_2 (q)$ (see Section 6.4 and
the paper [11]).

\noindent
THEOREM 3.1. {\it (i) A Hermitian IR with $c_1 \ge 0$ of the Poincar\'e
algebra ${\cal P}$ can be lifted to a Hermitian IR of the quantum
Poincar\'e algebra $U_q ({\cal P})$. \hfill\break
(ii) A Hermitian IR with $c_1(q)\ge 0$ of $U_q({\cal P})$ in the limit
$q \rightarrow 1$ becomes a Hermitian IR of the usual Poincar\'e algebra
${\cal P}$.}

\noindent
{\it Sketch of proof}. The validity of this theorem follows immediately
from the results of the paper [11] where explicit formulas of
a connection between the generators of the quantized and non-quantized
Poincar\'e algebras was found.

\noindent
4. ADJOINT ACTION AND TENSOR OPERATOR FOR QPA

Now we would like to recall the general definition of the adjoint action
for an arbitrary Hopf algebra ${\cal A}=(A,\Delta,S,\varepsilon)$.

Let $X$ be an $A$-bimodule, i.e. $X$ is a linear space of two
representations of $A$ which act on the left-side and on the right-side
of elements $x \in X$.
Then the adjoint action "{\rm ad}" is a homomorphism
$A \to {\rm End}\ X$ defined by the formula~:
$$
{\rm ad}_a x = (({\rm id} \otimes S) \Delta (a)) \circ x
\eqno (4.1)
$$
for any $x \in X$ and any $a \in A$, where the symbol "$\circ$" is
defined by the rule
$$
(a \otimes b) \circ x = a x b \ .
\eqno (4.2)
$$
If $\Delta (a) = \sum_i a_i \otimes b_i$ then an explicit form of the
adjoint action is
$$
{\rm ad}_a x = \sum_i a_i x S (b_i ) \ .
\eqno (4.3)
$$
If the linear space $X$ is irreducible with respect to the adjoint
action then the representation is called irreducible too.

\noindent
{\it Remark}. As the Hopf algebra $A$ is an $A$-bimodule, by
definition, then we have the canonical actions ${\rm ad}_A$ on $A$.

Let us show that the notation of the adjoint action is natural
in theoretical physics where the representation theory is used.
Indeed, let $R$ be a representation space of any Lie algebra ${\cal L}$
or its universal enveloping algebra $U({\cal L} )$.
Since the representation of ${\cal L}$ in $X$ is a homomorphism
${\cal L} \to {\rm End}\ R$ we have some left-side and right-side
actions of ${\cal L}$ in ${\rm End}\ R$, i.e. the representation
${\cal L} \to {\rm End}\ R$ induces simultaneously an
${\cal L}$-bimodule structure in the linear space ${\rm End}\ R$.
For the case of a usual Lie algebra ${\cal L}$ the adjoint action is
reduced to the usual commutator, i.e.
$$
{\rm ad}_a x = [a,x]
\eqno (4.4)
$$
for any $x \in {\rm End}\ R$ and any $a \in {\cal L}$.

If an $A$-bimodule $X$, where $A$ is an arbitrary Hopf algebra,
is a space of linear operators then we shall call $X$ a space of
tensor operators.
Therefore the words "tensor operators" mean that there is an
adjoint action of the Hopf algebra $A$ in the operator space $X$.
If the representation of the adjoint action is irreducible then
the tensor operator is called irreducible too.

Now we write down the explicit formulas for the adjoint action of
the quantum Poincar\'e algebra $U_q ({\cal P} )$~:
$$
{\rm ad}^{(q')}_{J_j} x = [J_j ,x]\  ,
\eqno {\rm (4.5a)}
$$
$$
{\rm ad}^{(q')}_{P_j}x=
[P_j ,x]q'^{-{P_0 \over 2}}-[q'^{-{P_0 \over 2}},x]P_j \ ,
\eqno {\rm (4.5b)}
$$
$$
{\rm ad}^{(q')}_{P_0} x = [P_0 ,x] \ ,
\eqno {\rm (4.5c)}
$$
$$
{\rm ad}^{(q')}_{N_j} x= [N_j ,x] q'^{-{P_0 \over 2}} -
[q'^{-{P_0 \over 2}},x]N_j + {3i \over 2} (\ln q')
[q'^{-{P_0 \over 2}},x]P_j \, -
$$
$$
-{1 \over 2} (\ln q')\varepsilon_{jlk} \Bigl( [P_l ,x]J_k
q'^{-{P_0 \over 2}} + q'^{-{P_0 \over 2}} [J_l ,x]P_k +
[q'^{-{P_0 \over 2}} ,x]J_l P_k \Bigr) \ ,
\eqno {\rm (4.5d)}
$$
for j = 1,2,3. If the four-momenta $P_{\mu}$, $(\mu = 0,1,2,3)$ commute
with $x$,
i.e. $[P_{\mu} ,x] = 0$, then the formulas (4.5a-d) take the form
$$
{\rm ad}^{(q')}_{P_{\mu}} x = 0\ ,
\eqno {\rm (4.6a)}
$$
$$
{\rm ad}^{(q')}_{J_j} x = [J_j ,x]\ ,
\eqno {\rm (4.6b)}
$$
$$
{\rm ad}^{(q')}_{N_j} x = [N_j ,x] q'^{-{P_0 \over 2}} - {1 \over 2}
(\ln q') \varepsilon_{jlk} q'^{-{P_0 \over 2}} [J_l ,x]P_k \ .
\eqno {\rm (4.6c)}
$$

Now we want to extend the representation "{\rm ad}" to a representation
of $U_q ({\cal P} )$ considerd as a Hopf algebra.
For the sake of convenience we shall use the following notations~:
$$
{\rm ad}^{(1)} := {\rm ad}^{(q)}\ , \ \ \
{\rm ad}^{(2)} := {\rm ad}^{(\bar q)}\ .
\eqno (4.7)
$$

\noindent
Let $X$ and $Y$ be two $U_q ({\cal P} )$-bimodules then we put
$$
{\rm ad}^{(s,l)}_{a\otimes b} (x \otimes y):=({\rm ad}^{(s)}_a \otimes
{\rm ad}^{(l)}_b)(x\otimes y)={\rm ad}^{(s)}_ax\otimes{\rm ad}^{(l)}_by
\eqno (4.8)
$$
for any $a,b \in U_q({\cal P})$, $x \in X$, $y \in Y$, and $s,l = 1,2$.
It is not difficult to show that the mapping $a \mapsto
{\rm ad}^{(s,l)}_{\Delta_{q'}(a)}$ (for $\forall a \in U_q({\cal P})$)
is a homomorphism. Therefore putting
$$
\Delta_{q'}({\rm ad}_a):={\rm ad}^{(s,l)}_{\Delta_{q'}(a)}
\eqno (4.9)
$$
for any $a\in U_q({\cal P})$ we define representations of the Hopf
algebra $(U_q({\cal P}), \Delta_{q'}$, $S_{q'} ,\varepsilon)$.

\noindent
The explicit expressions of the adjoint actions (4.9) on any element
$x \otimes y$ have the forms~:
$$
{\rm ad}^{(s,l)}_{\Delta_q(a)}(x \otimes y)=
\sum_i{\rm ad}^{(s)}_{a_i} x\otimes {\rm ad}^{(l)}_{b_i} y \ ,
\eqno (4.10)
$$
$$
{\rm ad}^{(s,l)}_{\Delta'_q(a)}(x\otimes y)=
\sum_i{\rm ad}^{(s)}_{b_i}x\otimes{\rm ad}^{(l)}_{a_i}y
\eqno (4.11)
$$
for any $a \in U_q({\cal P})$, where $\Delta_q(a)=\sum_i a_i\otimes b_i$
and $s,l = 1,2$.

\noindent
{\it Remark}. The formulas (4.10) and (4.11) define representations for
any Hopf algebra $(A ,\Delta ,S,\varepsilon )$ [12].

Now let the $U_q({\cal P})$-bimodule $X$ be an associative algebra, then
we can consider the actions of ${\rm ad}^{(s)}$, $s=1,2$, on the product
of two elements $x$ and $y$ from $X$.
It is not difficult to verify that the following formulas are valid
$$
{\rm ad}^{(1)}_axy=
\sum_i{\rm ad}^{(1)}_{a_i}x \ {\rm ad}^{(1)}_{b_i} y \ ,
\eqno (4.12)
$$
$$
{\rm ad}^{(2)}_axy=
\sum_i{\rm ad}^{(2)}_{b_i} x \ {\rm ad}^{(2)}_{a_i} y \ ,
\eqno (4.13)
$$
if $\Delta_q(a)=\sum_i a_i \otimes b_i$.
These formulas say that the multiplication of any two elements $x$ and
$y$ from $X$ is transformed with respect to the adjoint actions
${\rm ad}^{(1)}$ and ${\rm ad}^{(2)}$ in the same way as the tensor
product $x \otimes y$ is transformed with respect to
${\rm ad}^{(11)}_{ \Delta_q}$ and ${\rm ad}^{(22)}_{\Delta'_q}$
correspondingly.

The formulas (4.10), (4.11) and also (4.12), (4.13) allow us to
construct new tensors by coupling two and more tensors.

\noindent
{\it Remark}. The formulas (4.12), (4.13) are valid for any Hopf
algebra $(A ,\Delta ,S,\varepsilon )$ [12].

\noindent
5. SCALARS AND INVARIANTS OF QPA

In the case of usual (non-quantized) Lie algebras, as a rule, the
terms "scalar" and "invariant" are used as synonyms.
In the quantum case, i.e. for noncocommutative Hopf algebras,
we cannot use these terms as synonyms because they correspond to
two different notions.
Let us remember the exact definitions of these notions for our case
$U_q({\cal P})$.

{(i)}{\it An element $x \in X$ is called a scalar (or a tensor of
zero rank) if it satisfies the equations}
$$
{\rm ad}^{(q')}_{g_s}x=0
\eqno (5.1)
$$
{\it for any $g_s \in g:=\{J_l, N_k, P_{\mu}\ \vert \ l,k = 1,2,3;\
\mu = 0,1,2,3 \}$}.

{(ii)} {\it An element $x \in X$ is called an invariant if it
satisfies the equation~}:
$$
[a,x] = 0
\eqno (5.2)
$$
\noindent
{\it for any $a \in U_q({\cal P})$}.

 From the relation (4.5 a-d) follows the simple theorem.

\noindent
THEOREM 5.1. {\it For some element $x \in X$ and for any
$a \in U_q({\cal P})$ and for any $g_s \in g$ we have}
$$
[a,x] = 0 \buildrel \rm (iff) \over \Longleftrightarrow
{\rm ad}^{(q')}_{g_s}x=0 \ .
\eqno (5.3)
$$

\noindent
{\it Remark}. (i) The Theorem 5.1 is valid for any quantized
contragredient
Kac-Moody (super) algebra (in a broad sense [13]) [12]. \hfill\break
(ii) For usual Lie algebras the property (5.3) is trivial (because,
for example, ${\rm ad}_{g_s} x = [g_s ,x])$.

\noindent
We shall call scalars and invariants of types (5.1) and (5.2)
one-particle types.
Now we want to consider some properties of scalars and invariants of
two-particle types.
The following theorem is valid.

\noindent
THEOREM 5.2. {\it (about two-particle scalars and invariants).
Let $X$ and $Y$ be two $U_q ({\cal P} )$-bimodules (the spaces of
tensor operators) and let $\sum_m x_m \otimes y_m$ be some element
of $X \otimes Y$ then we have}
$$
{\rm ad}^{(2,1)}_{\Delta_q(g_s)}(\sum_m x_m \otimes y_m)=0
\buildrel \rm (iff) \over \Longleftrightarrow [\Delta_q(g_s),
\sum_m x_m \otimes y_m] = 0
\eqno (5.4)
$$
{\it and also}
$$
{\rm ad}^{(1,2)}_{\Delta'_q (g_s)}(\sum_m x_m \otimes y_m)=0
\buildrel \rm (iff) \over \Longleftrightarrow [\Delta'_q(g_s),
\sum_m x_m \otimes y_m]=0
\eqno (5.5)
$$
{\it simultaneously for all elements $g_s \in g := \{ J_l ,N_k
,P_{\mu} \ \vert \ l,k = 1,2,3; \mu = 0,1,2,3 \}$}.

\noindent
{\it Proof}. We prove the first part of the theorem (i.e. the property
(5.4)) the second part (5.5) is proven in a similar way.
First of all it is evident that
$$
[\Delta_q (J_j),\sum_m x_m \otimes y_m]\equiv {\rm ad}_{\Delta_q(J_j)}
(\sum_m x_m \otimes y_m ), \
\eqno (5.6)
$$
$$
[\Delta_q(P_0),\sum_m x_m \otimes y_m] \equiv {\rm ad}_{\Delta_q(P_0)}
(\sum x_m \otimes y_m)\ .
\eqno (5.7)
$$
Further it is easy to show that if $\Delta_q (P_0 )$ commutes with
$\sum_m x_m \otimes y_m$ then we have
$$
[\Delta_q (P_j ),\sum_m x_m \otimes y_m] =
\Bigl({\rm ad}^{(2,1)}_{\Delta_q(P_j)}(\sum_m x_m \otimes y_m)\Bigr)
(q^{-{P_0 \over 2}} \otimes q^{{P_0 \over 2}}) \ .
\eqno (5.8)
$$
Analogously but by more cumbersome calculation we obtain that if the
generators $\Delta_q (J_j )$, $(j = 1,2,3)$, $\Delta_q(P_{\mu})$,
$(\mu = 0,1,2,3)$, commute with the tensor $\sum_m x_m \otimes y_m$
then
$$
[\Delta_q (N_j ),\sum_m x_m \otimes y_m ] =
\Bigl({\rm ad}^{(2,1)}_{\Delta_q(N_j)}(\sum_m x_m \otimes y_m)\Bigr)
(q^{-{P_0 \over 2}} \otimes q^{{P_0 \over 2}}) \ .
\eqno (5.9)
$$
Thus if the left hand sides of the relations (5.6)-(5.9) are
equal to zero simultaneously then the right hand sides of the
relations are equal to zero too and vice versa.
The second part (5.5) of the theorem is proven in a similar way.

\noindent
{\it Remark}. An analog of such a theorem is also valid for any
quantized Kac-Moody (super) algebra [12].

Theorem 5.2 about two-particle scalars and invariants and also
Theorem 5.1 for the one-particle case are very important for the
constructions of q-relativistic equations, i.e. the q-analogs of
the relativistic equations which are invariant under the quantum
Poincar\'e algebra $U_q (P)$.
In Section 7 we shall demonstrate this in explicit form.
Theorem 5.2 is also important for the construction of the Casimir
operators for many-particle systems.

\noindent
6. EXPLICIT FORMS OF SOME IMPORTANT IRREDUCIBLE TENSOR
OPERATORS FOR QPA

In this section we find explicit forms of three covariant four-vectors
for the quantum Poincar\'e algebra $U_q ({\cal P} )$, i.e. we construct
three four-dimensional irreducible representations with respect to the
adjoint action of $U_q ({\cal P} )$.
The first representation is realized in the space $U_q (P_0 ,\vec P)$
of smooth functions of the four-momenta $P_{\mu}$, $\mu = 0,1,2,3$.
The second representation is realized in the space $U_q (P_0 ,\vec P,
W_0 ,\vec W)$ which is an universal enveloping algebra generated by
the four-momenta and the Pauli-Lubanski four-vector of "classical"
type.
The third representation is realized in the space of the usual
Dirac $\gamma$-matrices.
The first two representations give us an answer to the question~:
what are the right (covariant) components of the four-momenta and
the Pauli-Lubanski four-vector. At the end of the section we construct
two Casimir operators of $U_q ({\cal P})$ by couplings of two covariant
four-momenta and also two covariant Pauli-Lubanski four-vectors of
"quantum" type. We start with the four-momenta.

\noindent
6.1. {\bf The covariant four-momenta}

The following simple proposition is valid.

\noindent
PROPOSITION 6.1. {\it Let $U(P_0 ,\vec P)$ be an universal enveloping
algebra (algebra of smooth functions) of the four-momentum sector of
$U_q ({\cal P} )$.
Then the linear space $U_q (P_0 ,\vec P)$ is invariant with
respect to the adjoint action of $U_q ({\cal P} )$ and this
representation is equivalent to a representation of the adjoint
action of ${\cal P}$ (or the Lorentz algebra ${\cal L}$) on the space
$U_{q=1} (P_0 ,\vec P)$}.

The proof follows immediatly from the relation (4.6a-c) and (2.1a-h).

\noindent
PROPOSITION 6.2. {\it The new four-momenta $P_{\mu} (q')$, $\mu =
0,1,2,3$, defined by
$$
P_j(q') = P_j q'^{{P_0 \over 2}}\ ,\ \ \  P_0 (q') = [P_0] +
{1 \over 2} (\ln q')\vec P^2
\eqno (6.1)
$$
for $j = 1,2,3$, are transformed with respect to the adjoint action
of $U_q ({\cal P} )$ exactly as the usual (non-deformed) four-momenta
are transformed with respect to the adjoint action of the
non-deformed Poincar\'e algebra ${\cal P}$}.

In the relation (6.1) and everywhere we use the nearly standard
notation~:
$$
[P_0]:={q'^{P_0}-q'^{-P_0} \over 2(\ln q')}={q^{P_0}-q^{-P_0}
\over 2(\ln q)} \ .
\eqno (6.2)
$$
The components (6.1) are naturally called the covariant four-momenta.

\noindent
{\it Proof}. Let $J_{\pm}:=J_1 \pm iJ_2$,\ \ \ $N_{\pm}:=N_1 \pm iN_2$
\ \ \ and
$$
K^{(s)}_{\pm}:=J_{\pm}-(-1)^s iN_{\pm}\ ,\ \ \
K^{(s)}_0:=J_3-(-1)^s iN_3
\eqno (6.3)
$$
for $s = 1,2$.
It is easy to verify that the operators ${\rm ad}^{(q')}_{K^{(1)}_j}$
and ${\rm ad}^{(q')}_{K^{(2)}_{j'}}$ commute one with the other in
the space $U_q (P_0 ,\vec P)$, i.e.
$$
[{\rm ad}^{(q')}_{K^{(1)}_j},\ {\rm ad}^{(q')}_{K^{(2)}_{j'}}] U_q
(P_0 ,\vec P) = 0
\eqno (6.4)
$$
for any $j,j' = +,-,0$.
The operators ${\rm ad}^{(q')}_{K^{(1)}_j}$ and
${\rm ad}^{(q')}_{K^{(2)}_{j'}}$ generate the usual complex Lorentz
algebra $so(4,\complex)=so(3,\complex)\oplus so(3,\complex)$.
Let
$$
T_{\pm ,\pm}(q')=\pm (P_1 \pm iP_2)q'^{{P_0 \over 2}} \ ,
$$
$$
T_{\pm ,\mp}(q')=-P_3 q'^{{P_0 \over 2}} \pm \Bigl([P_0]+
{1 \over 2} (\ln q'){\vec P}^2 \Bigr)\ .
\eqno (6.5)
$$
Then the proof of the Proposition 6.2 is reduced to checking of
the following relations
$$
{\rm ad}^{(q')}_{K^{(1)}_{\pm}} T_{\pm ,j}(q')=0 \ , \ \ \
{\rm ad}^{(q')}_{K^{(2)}_{\pm}} T_{j,\pm} (q')=0 \ ,
$$
$$
{\rm ad}^{(q')}_{K^{(1)}_0} T_{\pm ,j}(q')=\pm{1\over 2}T_{\pm,j}(q')\ ,
\ \ \ {\rm ad}^{(q')}_{K^{(2)}_0} T_{j,\pm} (q')=\pm
{1 \over 2} T_{j,\pm} (q') \ ,
\eqno (6.6)
$$
$$
{\rm ad}^{(q')}_{K^{(1)}_{\pm}} T_{\mp ,j}(q')=T_{\pm ,j} (q') \ ,\ \ \
{\rm ad}^{(q')}_{K^{(2)}_{\pm}} T_{j,\mp}(q')=T_{j,\pm} (q') \ ,
$$
for $j = +,-$.
This is easy to verify by the formulas (4.6a-c).

\noindent
6.2. {\bf The covariant Pauli-Lubanski four-vector}

Some components of the q-Pauli-Lubanski four-vector were suggested in
the papers [3,4]. However these components are not covariant because
they have not the right transformation properties with respect to
the adjoint action of $U_q ({\cal P} )$.
The covariant components of the Pauli-Lubanski four-vector are
described by the following proposition:

\noindent
PROPOSITION 6.3. {\it The four-vector $W_{\mu} (q')$, $\mu=0,1,2,3$,
defined by
$$
W_j (q')=J_j [P_0]+\varepsilon_{jlk} P_l N_k +
{1 \over 2} (\ln q') P_j (\vec P \vec J) \ ,
$$
$$
W_0 (q') = (\vec P \vec J) q'^{{P_0 \over 2}}
\eqno (6.7)
$$
for $j=1,2,3$ is transformed with respect to the adjoint action of
$U_q({\cal P})$ exactly as the usual (non-deformed) Pauli-Lubanski
four-vector is transformed with respect to the adjoint action of
the non-deformed Poincar\'e algebra ${\cal P}$}.

\noindent
It is naturally called the covariant Pauli-Lubanski four-vector.
The Pauli-Lubanski four-vector introduced in [3,4] differs from (6.7)
by the factor $q'^{{P_0 \over 2}}$ and by the additional term
${1 \over 2} (\ln q') \vec P (\vec P \vec J)$.

\noindent
{\it Proof}. First of all we verify that the operators
${\rm ad}^{(q')}_{P_{\mu}}$, $\mu = 0,1,2,3$ act as the zero operator,
i.e. ${\rm ad}^{(q')}_{P_{\mu}}W_\nu(q')=0$, for $\mu, \nu=0,1,2,3$.
Further the proof is reduced to checking the relations (6.6) where
the components $T_{j,j'}(q')$, $(j,j'=+,-)$, have now the form
$$
T_{\pm ,\pm} (q'):=(W_1 (q') \pm W_2(q')) \ ,
$$
$$
T_{\pm ,\mp} (q'):=-(W_3 (q') \pm W_0 (q')) \ .
\eqno (6.8)
$$

It is well-known that an algebra generated by components of the usual
Pauli-Lubanski four-vector plays an important role for
the classification of the irreducible representations of
the Poincar\'e algebra ${\cal P}$.
The same is valid for $U_q ({\cal P})$.
Therefore we write down the commutation relations for the components of
the q-analog Pauli-Lubanski four-vector $W_{\mu} (q')$, $\mu =
0,1,2,3$~:
$$
[W_j (q') ,W_l (q')]= i\varepsilon_{jlk} (P_0 (q')
W_k (q')-W_0 (q') P_k(q')) \ ,
$$
$$
[W_0 (q'),W_j(q')]=-i\varepsilon_{jlk} P_l(q')W_k(q')
\eqno (6.9)
$$
and also
$$
[P_{\mu} (q'), W_{\nu}(q')]=0\ ,
\eqno (6.10)
$$
$$
P_{\mu} (q') W^{\mu}(q'):=g^{\mu\nu} P_{\mu}(q')W_{\nu}(q')=0\ ,
\eqno (6.11)
$$
where $g_{\mu\nu}$ is the metric tensor~:
$$
g_{\mu\nu}=g^{\mu\nu}:=\ {\rm diag}\ (1,-1,-1-1)\ .
\eqno (6.12)
$$

\noindent
6.3. {\bf Realization of $U_q({\cal P})$ by the usual Dirac
$\gamma$-matrices}

It is well-known that the usual Lorentz algebra has a four-dimensional
realization in terms of the Dirac $\gamma$-matrices $\gamma_{\mu}$,
$\mu = 0,1,2,3$, which have the following basic property
$$
\{\gamma_{\mu}, \gamma_{\nu}\}:=\gamma_{\mu} \gamma_{\nu} +
\gamma_{\nu} \gamma_{\mu}=2g_{\mu\nu} \ .
\eqno (6.13)
$$
This representation can be considered simultaneously as a representation
of ${\cal P}$ and $U_q({\cal P})$, for which the four-momenta generators
are realized as the zero operator. For the sake of convenience we write
down this realization in the explicit form
$$
\hat J_j={1 \over 4} i\varepsilon_{jlk} \gamma_l \gamma_k \ ,
\eqno {\rm (6.14a)}
$$
$$
\hat N_j={1\over 2} \gamma_0 \gamma_j \ ,
\eqno {\rm (6.14b)}
$$
$$
\hat P_{\mu} = 0 \ ,
\eqno {\rm (6.14c)}
$$
for $j = 1,2,3$ and $\mu = 0,1,2,3.$

\noindent
The $\gamma$-matrices $\gamma_{\mu}$, $\mu=0,1,2,3,$ sandwiched between
Dirac spinors, are transformed with respect to the adjoint actions of
the operators (6.14a-c) exactly as the covariant components
$P_{\mu} (q')$ (or $W_{\mu} (q')$) are transformed with respect to
${\rm ad}_{J_j}$ and ${\rm ad}_{N_j}$, $j = 1,2,3$.

\noindent
6.4. {\bf The Casimir operators}

Using the covariant tensor components of the four-momenta and the
Pauli-Lubanski four-vector we can easily construct the Casimir
operators of $U_q ({\cal P} )$ by scalar coupling in accordance with the
relations (4.12), (4.13) and Theorem 5.1.

The first scalar invariant $I_1 (q')$ is constructed by the
coupling of two four-momenta~:
$$
I_1 (q')=P_{\mu}(q') P^{\mu}(q')=\Bigl([P_0] + {1 \over 2}
(\ln q')\vec P^2 \Bigr)^2 - \vec P^2 q'^{P_0}
\eqno (6.15)
$$
where $[P_0 ] = {1 \over 2} (\ln q')^{-1} (q^{P_0} -q^{-P_0} )$.
After simple manipulations we obtain that
$$
I_1 (q') = C_1 (q')\Bigl(1 + {1 \over 4} (\ln q')^2 C_1(q') \Bigr)
\eqno (6.16)
$$
where
$$
C_1 (q') = 4[{P_0 \over 2}]^2 - \vec P^2 \ .
\eqno (6.17)
$$
However we can easily verify that the operator $C_1 (q')$ is an
invariant too [2].
Moreover this operator cannot be represented as a coupling of some
tensor operators.
This property is typical for all quantized simple Lie algebras (see,
for example, $U_q (sl(2,\complex ))$ [14-16]).
The operator $C_1 (q')$ is naturally called the first Casimir operator
of $U_q ({\cal P} )$.

The second scalar invariant $I_2 (q')$ is obtained by coupling two
Pauli-Lubanski four-vectors~:
$$
I_2 (q') = W_{\mu} (q') W^{\mu} (q')\ .
\eqno (6.18)
$$
After a simple manipulation, the invariant takes the form
$$
I_2(q')=\Bigl({q^{P_0}+q^{-P_0} \over 2}-{1 \over 4}(\ln q')^2
\vec P^2 \Bigr)(\vec P\vec J)-\Bigl(\vec J [P_0]+
\vec P \times \vec N\Bigr)^2\ .
\eqno (6.19)
$$
This expression is well-known [3,4] and is called the second Casimir
operator of $U_q({\cal P})$.
Thus we see that the second Casimir operator of $U_q({\cal P})$ is
exactly given by the scalar coupling of two covariant Pauli-Lubanski
four-vectors. This Casimir operator is usually denoted by $C_2(q')$.

It should be noted that both scalar invariants $I_1(q')$, $I_2(q')$ and
also the first Casimir operator $C_1 (q')$ do not change if we put
$q'=q$ or $\bar q$. Therefore we can remove the index prime of
the argument $q'$.

\noindent
7. CONSTRUCTION OF q-ANALOGS FOR SOME RELATIVISTIC EQUATIONS

For the construction of q-analogs of relativistic equations we need
some minimal (spinless) differential realization of $U_q({\cal P})$ on
the Minkovski space-time manifold. Such a realization was already
considered in Refs [3,4] and has the following form~:
$$
P_{\mu}=-i\partial_{\mu}\ ,\ \
J_j=-i\varepsilon_{jlk} x^l \partial_k\ , \ \
N_j=-ix_0 \partial_j + x_j [-i\partial_0])
\eqno (7.1)
$$
where $\partial_{\mu} := \partial /\partial x^{\mu}$.

We start our consideration with the Klein-Gordon equation.
For a construction of a q-analog of this equation there exist two
possibilities~: either to use the first Casimir operator $C_1 (q)$
(6.17) or to use the invariant $I_1 (q)$ (6.16).
Thus we have two q-analog variants of the Klein-Gordon equation,
namely
$$
(4[{P_0 \over 2}]^2 - \vec P^2)\Psi = 4[{m \over 2}]^2 \Psi
\eqno (7.2)
$$
and
$$
P_{\mu} (q)P^{\mu} (q)\Psi = 4[{m \over 2}]^2 \Bigl(1 + (\ln q)^2
[{m \over 2}]^2 \Bigr)\Psi \ .
\eqno (7.3)
$$
We want to stress that in the equation (7.2) (and also (7.3)) we use
a connection of the Casimir operator eigenvalue with the mass parameter
$m$ given by the q-number (6.2). In the literature up to now the mass
parameter $m$ on the right hand side of (7.2) was used without brackets.
There are at least two reasons for our choice~:

\begin{description}
\item{(i)} The mass $m$ does not depend on $q$ when three momentum
components are equal to zero, $p_1 = p_2 = p_3 = 0$, because such
a representation of $U_q({\cal P})$ coincides with a representation of
the usual Poincar\'e algebra (see [17]).

\item{(ii)} There exists a similar connection between the highest
weight and the Casimir operator eigenvalue for IR's as is the case for
quantized simple Lie algebras (see [14-16] for example, for
$U_q(sl(2,C))$.
\end{description}

Now we want to construct a q-analog of the Dirac operator.
This q-analog has to be an invariant of a representation space for
the tensor product representations ${\rm ad}_{\hat g_s}$ and
${\rm ad}^{(q)}_{g_s}$ where $\hat g_s$ are the operators of the
$\gamma$-matrix realization (6.14a-c) and $g_s$ are the operators of
the differential realization (7.1).
A basis of this representation space consists of the elements
$\gamma_{\mu} \otimes P^{\nu} (q)$, $\mu ,\nu = 0,1,2,3$.
Since matrix elements of the $\gamma$-matrices commute with the
components of the four-momenta, we may remove here the symbol of
the tensor product, i.e. we put
$$
\gamma_{\mu} P^{\nu} (q) := \gamma_{\mu} \otimes P^{\nu} (q) \ ,
\eqno (7.4)
$$
for $\mu,\nu = 0,1,2,3$. It should be noted that the $\gamma$-matrix
basis does not depend on the parameter $q$.
In accordance with Theorem 5.2 (about two-particle scalars and
invariants) we have the q-analog of the Dirac operator
$$
D(q):=\gamma_{\mu} P^{\mu}(q)=\gamma_0 \Bigl([P_0]+
{1 \over 2}(\ln q)\vec P^2 \Bigr)-\vec \gamma \vec P q^{{P_0 \over 2}} \ .
\eqno (7.5)
$$
It is not difficult to verify that the square of this operator is
proportional to the invariant (6.15). More exactly we have
$$
D^2(q)=I_1 (q) \ .
\eqno (7.6)
$$
We write down the explicit form of the global generators which commute
with the q-Dirac operator (7.5)~:
$$
{\cal J}_j = \hat J_j + J_j \ ,
\eqno {\rm (7.7a)}
$$
$$
{\cal N}_j=\hat N_j q^{{P_0 \over 2}}+N_j+
{1 \over 2}(\ln q) \varepsilon_{jlk} \hat J_l P_k \ ,
\eqno {\rm (7.7b)}
$$
$$
{\cal P}_{\mu} = P_{\mu}
\eqno {\rm (7.7c)}
$$
where $\hat J_j$, $\hat N_j,  (j = 1,2,3)$ are the operators of the
$\gamma$-matrix realization (6.14a-c) and $J_j$, $N_j$, $(j=1,2,3)$,
$P_{\mu}, (\mu = 0,1,2,3)$, are the operators of the differential
realization (7.1). The expressions (7.7a-c) are obtained in
the following simple way. We take the formulas for the coproduct
(2.2a-d) and put the $\gamma$-matrix realization instead of
the first components and the differential realization instead of
the second components.

\noindent
At last we write the q-analog of the Dirac equation
$$
\Bigl(\gamma_0([P_0]+{1 \over 2}(\ln q)\vec P^2)-
\vec \gamma \vec P q^{{P_0 \over 2}} \Bigr)\Psi =
2[{m \over 2}](1+(\ln q)^2 [{m \over 2}]^2)^{1/2} \Psi \ .
\eqno (7.8)
$$
The dependence of the right hand side on the mass parameter $m$ is
defined by the demand that the square of the expression coincides
with the right hand side of the equation (7.3). The q-Dirac equation was
given in ref [18,7] with ${m \over 2}$ instead of $[{m \over 2}]$, but
the derivation given here is much simpler.

Thus we have shown that the right covariant components of the tensor
operator give powerfool tools for the construction of q-relativistic
equations. In such a way we can construct the q-relativistic equations
for spin $s = 1,3/2$ and other equations.
We believe a q-analog of the Maxwell equation has the form
$$
P_{\mu} (q')P^{\mu} (q')A_{\nu} - P_{\nu} (q')P_{\mu} (q')A^{\mu} =
j_{\nu} \ .
\eqno (7.9)
$$
We shall discuss this problem in detail in another paper.

\noindent
8. THE WIGNER-ECKART THEOREM FOR QPA

As is well known the Wigner-Eckart theorem plays a very important
role for calculations of quantum mechanical one- and many-body
systems, where group representation theory is used.
We believe that this theorem is important also for an arbitrary
Hopf algebra \footnote
{This theorem for the case of
$U_q (sl(2,\complex))$ was considered extensively in Refs [15,16]
and for an arbitrary quantized simple Lie algebra in [14].}
in particularly for $U_q({\cal P})$.

Let us come back again to an arbitrary Hopf algebra ${\cal A} =
(A,\Delta ,S,\varepsilon )$ (see Section 4).
Let $V$ be a vector space of a representation of a Hopf algebra
$A$ and also its $A$-bimodule $X$, i.e. we have two homomorphisms
$A \to {\rm End}\ V$ and $X \to {\rm End}\ V$ simultaneously.
Let us consider vectors of the form $ax \vert\psi\rangle$,
where $a\in A$, $x \in X$, $\vert\psi\rangle \in V$.

\noindent
PROPOSITION 8.1. {\it If $\Delta (a) = \sum_i a_i \otimes b_i$ then
the following relations are valid
$$
ax\vert\psi\rangle=\sum_i({\rm ad}^{(1)}_{a_i} x)b_i \vert\psi\rangle
\eqno (8.1)
$$
and}
$$
ax\vert\psi\rangle=\sum_i({\rm ad}^{(2)}_{b_i} x)a_i \vert\psi\rangle
\eqno (8.2)
$$

\noindent
{\it Proof}. Writing the adjoint actions ${\rm ad}^{(1)}_{a_i}$ and
${\rm ad}^{(2)}_{b_i}$ for the right parts of (8.1) and (8.2) in
the explicit form and using standard properties of the coproduct
$\Delta$ and the antipode $S$ we obtain immediately the left parts of
the relations (8.1) and (8.2).

The formulas (8.1) and (8.2) affirm that the vector $x\vert\psi
\rangle$ is transformed by the element $a\in A$ exactly the same
way as the vector $x \otimes \vert\psi\rangle$ is transformed by
the operators $({\rm ad}^{(1)} \otimes {\rm id})\Delta (a) =
\sum_i{\rm ad}^{(1)}_{a_i} \otimes b_i$ and
$({\rm ad}^{(2)} \otimes {\rm id})\Delta '(a)=
\sum_i {\rm ad}^{(2)}_{b_i} \otimes a_i$, i.e. the vector space $XV$
is transformed with respect to the Hopf algebra $A$ exactly as
the vector space $X \otimes V$ is transformed with respect to
$\Delta (A)$ and $\Delta '(A)$, where the first components of
the coproduct act by adjoint action on $X$.

This proposition is the basis for the proof of the Wigner-Eckart
theorem for an arbitrary Hopf algebra $A$.
Let us formulate and prove this theorem for $U_q({\cal P})$.
Let
$$
P_0, P_1, P_2, P_3 ,(\vec P \vec J),C_1 (q),C_2 (q)
\eqno (8.3)
$$
be a complete set of commuting operators.
Let $\{ \vert c,\tilde p,\lambda\rangle \}$ be an orthogonal basis
of some Hermitian IR, where $c = (c_1, c_2)$ are parameters connected
with eigenvalues of the Casimir operators $(C_1 (q), C_2 (q))$;
$\tilde p=(p_0,p_1,p_2,p_3)$ are eigenvalues of the operators $P_{\mu}$,
$\mu = 0,1,2,3$, and $\lambda$ is a eigenvalue of the operator
$(\vec P \vec J)$, i.e.
$$
C_{s}(q)\vert c,\tilde p,\lambda\rangle=c_{s}\vert c,
\tilde p,\lambda \rangle \ ,
\eqno {\rm (8.4a)}
$$
$$
P_{\mu} \vert c,\tilde p,\lambda\rangle = p_{\mu} \vert c,
\tilde p,\lambda\rangle \ ,
\eqno {\rm (8.4b)}
$$
$$
(\vec P \vec J)\vert c,\tilde p,\lambda \rangle = \lambda p \vert
c,\tilde p ,\lambda\rangle\ ,\ p = \sqrt{{\vec p}\ ^2} \ .
\eqno {\rm (8.4c)}
$$
Further let $\{ x^{c'}_{\tilde p', \lambda '} \}$ be an IR basis
in the tensor operator space $X$.
The basis element $x^{c'}_{\tilde p ',\lambda '}$ is called
a component of the tensor operator.

\noindent
WIGNER-ECKART THEOREM. {\it The matrix element $\langle c'',\tilde p'',
\lambda ''\vert x^{c'}_{\tilde p',\lambda '} \vert c,\tilde p,
\lambda\rangle$ is proportional to a Clebsch-Gordan coefficient with
a constant which does not depend on intrinsic quantum number
$(p,\lambda ;p' ,\lambda';p'',\lambda '')$, i.e.
$$
\langle c'',\tilde p'',\lambda ''\vert x^{c'}_{\tilde p ',\lambda'}
\vert c,\tilde p,\lambda\rangle =
(c,\tilde p,\lambda ;c',\tilde p',\lambda'\vert c'',
\tilde p'',\lambda'')_q(c''\Vert x^{c'}\Vert c)_q
\eqno (8.5)
$$
where $(\dots ;\dots \vert \dots )_q$ is a Clebsch-Gordan coefficient
of $U_q ({\cal P} )$; $(c'\Vert x^{c'}\Vert c)_q$ is called the reduced
matrix element which depends on the quantum numbers $c,c'$ and $c''$
only.}

\noindent
{\it Remark}. For the sake of simplicity we consider here in (8.5)
a multiplicity free case, i.e. when the vector
$x^{c'}_{\tilde p',\lambda '} \vert c,\tilde p,\lambda\rangle$
belongs to some Hilbert space, the direct sum of subspaces carrying
IRs of $U_q ({\cal P} )$ and each equivalence class of IRs occurs once
and only once.

\noindent
{\it Proof}. As follows from the Proposition 8.1 the vector
$x^{c'}_{\tilde p',\lambda '} \vert c,\tilde p,\lambda\rangle$
is transformed by an element $a\in U_q ({\cal P} )$ exactly as the
vector $x^{c'}_{\tilde p',\lambda'} \otimes \vert c,\tilde p,
\lambda\rangle$ is transformed by the representation
$\Delta_q (U_q ({\cal P} ))$, where the first components of this
coproduct act by adjoint action on $x^{c'}_{\tilde p',\lambda '}$.
Therefore the vector of the form
$$
\sum_{\lambda ,\lambda '} \int (c,\tilde p,\lambda ;c',\tilde p',
\lambda '\vert c'',\tilde p'',\lambda '')_q
x^{c'}_{\tilde p',\lambda'} \vert c,\tilde p,\lambda \rangle
d\tilde p d\tilde p'
\eqno (8.6)
$$
is transformed by $\Delta_q (U_q ({\cal P} ))$ as the IR $c''$, i.e.
the vector (8.6) is proportional to the vector $\vert c'',
\tilde p'',\lambda ''\rangle$~:
$$
\sum_{\lambda ,\lambda '} \int (c,\tilde p,\lambda ;
c',\tilde p',\lambda '\vert c'', \tilde p'',\lambda'')_qx^{c'}_{\tilde
p', \lambda '} \vert c,
\tilde p,\lambda\rangle d\tilde p d\tilde p'= \\
(c''\Vert x^{c'} \Vert c)_q\vert c'',\tilde p'', \lambda ''\rangle
\eqno (8.7)
$$
where the coefficient $(c''\Vert x^c \Vert c)_q$ does not depend on
the quantum numbers $\tilde p,\lambda ,\tilde p',\lambda ',
\tilde p'',\lambda ''$.
Converting back this relation we obtain
$$
x^{c'}_{\tilde p',\lambda '} \vert c,\tilde p,\lambda\rangle=
(c''\Vert x^{c'}\Vert c)_q \sum_{\lambda''',c'''}\int (c,\tilde p,
\lambda ;c',\tilde p',\lambda'\vert c''',\tilde p''',\lambda ''')_q
\vert c''' \tilde p'''\lambda '''\rangle d\tilde p''' \ .
\eqno (8.8)
$$
Taking matrix elements with the bra-vector $\langle c'',\tilde p'',
\lambda'' \vert$ we obtain the formula (8.5).

\noindent
9. CONCLUSION

In this paper we considered only a part of the problems of the
representation theory for the quantized Poincar\'e algebra.
In this conclusion we want to elucidate the program of our recent
and near future research~:
\begin{description}
\item[(i)] Develop the theory of induced representations for
$U_q ({\cal P} ))$.

\item[(ii)] Construct the explicit bases for the most important
(from the point of view of q-relativistic physics) irreducible
representations.

\item[(iii)] Apply the tensor operator theory to the construction
of q-relativistic kinematics.

\item[(iv)] Develop the representation theory for the quantized
super Poincar\'e and the Poincar\'e algebra in higher dimensions.

\item[(v)] Try to construct the q-analog of quantum electrodynamics.
\end{description}
\noindent
ACKNOWLEDGEMENTS

The second author (V.N.T.) would like to thank the University of
Geneva for its hospitality and the Swiss National Science Foundation
for its financial support.

\noindent
REFERENCES
\begin{enumerate}

\ref Celeghini, E. Giachetti, R., Sorace, E., and Tarlini, M.,
{\it J. Math. Phys.} {\bf 31}, 2548-2551 (1990); {\bf 32}, 1155-1158
(1991);{\bf 32}, 1159-1165 (1991) and "Contractions of quantum groups",
in Quantum Groups, Lecture Notes in Mathematics No 1510, 221
(Springer-Verlag, 1992).
\ref Lukierski, J., Nowicki, A., Ruegg, H., and Tolstoy, V.N.,
{\it Phys. Lett.} {\bf B264}, 331-338 (1991).
\ref Giller, S., Kunz, Y., Kosinski, P., Majewski, M., and
Maslanka, P., {\it Phys. Lett.} {\bf B286}, 57-61 (1992).
\ref Lukierski, J., Nowicki, A., and Ruegg, H., {\it Phys. Lett.}
{\bf B293}, 344-352 (1992).
\ref Lukierski, J., and Ruegg, H., "Quantum $\kappa$-Poincar\'e in any
Dimension", University of Geneva Preprint, UGVA-DPT 1993/08-825.
\ref Domokos, G., and Kovesi-Domokos, S., "Astrophysical Limit on the
Deformation of the Poincar\'e Group", Preprint JHK-TIPAC-920027/REV,
May 1993.
\ref Lukierski, J., Ruegg, H., and R\"uhl, W.,
{\it Phys. Lett.} {\bf B313}, 357-366 (1993).
\ref Biedenharn, L.C., Mueller, B., and Tarlini, M., "The Dirac-Coulomb
Problems for the $\kappa$-Poincar\'e Quantum Group", Universities of
Texas at Austin, Duke and Firenze, Preprint 1993.
\ref Ruegg, H., "q-Deformation of Semisimple and Non-Semisimple Lie
Algebras",{\it Integ\-rable Systems, Quantum Groups and Quantum Field
Theories}, Ibort, L.A. and Rodriguez, M.A., ed., Kluwer Academic
Publishers Dordrecht (1993) pp. 45-81.
\ref Giller, S., Gonera, C., Kosinski, P., Majewski M., Maslanka, P. and
Kunz, J. "On $q$-Covariant Wave Functions",  Mod. Phys. Lett. {\bf A40}
3785-3797 (1993).
\ref Maslanka, P., "Deformation Map and Hermitean Representations of
$\kappa$-Poincar\'e Algebra", J. Math. Phys. {\bf 34}, 6025-6029 (1993) .
\hfill\break
"Global Counterpart of $\kappa$-Poincar\'e Algebra and Covariant Wave
Functions", Preprint 5/93, IM UL.
\ref Tolstoy, V.N., "Tensor Operators for Quantized Kac-Moody (super)
Algebras, in preparation.
\ref Tolstoy V.N. "Extremal projectors for quantized Kac-Moody
(super) Algebras and some of their applications". The Proc. of
the Quantum Groups Workshop, Clausthal, Germany (July 1989).
{\it Lecture Notes in Physics } {\bf 370}, 118-125 (1990).
\ref Biedenharn, L.C., and Tarlini, M., {\it Lett. Math. Phys.}
{\bf 20}, 271-278 (1990).
\ref Smirnov, Yu.F., Tolstoy, V.N., and Kharitonov, Yu.I.,
{\it J. Nucl. Phys.} {\bf 53},No.4, 959-980 (1991); {\bf 53},No.6,
1746-1771 (1991); {\bf 55},No.10, 2863-2874 (1992); {\bf 56},No.5,
236-257 (1993) (in Russian).
\ref Rittenberg, V., and Scheunert, M., {\it J. Math. Phys.} {\bf 33},
436-445 (1992).
\ref Bacry, H., {\it Phys. Lett.} {\bf B306}, 41-43 (1993).
\ref Nowicki, A., Sorace, E., and Tarlini, M., {\it Phys. Lett.}
{\bf B302}, 419-422 (1993). \hfill\break

\end{enumerate}
\end{document}